\documentclass[a4paper,12pt]{article}
\usepackage{amsfonts}
\usepackage{latexsym}
\usepackage{amsmath}
\usepackage{amssymb}
\usepackage{amssymb}
\usepackage{slashed}
\usepackage{upgreek}
\usepackage{mathrsfs}
\usepackage{bbm}

\usepackage{amsthm}

\theoremstyle{remark}

\usepackage[utf8]{inputenc}
\usepackage[english]{babel}

\theoremstyle{definition}

\allowdisplaybreaks

\usepackage{relsize}
\usepackage{graphicx}
\usepackage{comment}

\renewcommand{\thefootnote}{\fnsymbol{footnote}}

\def\appendix#1{\addtocounter{section}{1}\setcounter{equation}{0}
\renewcommand{\thesection}{\Alph{section}}
\section*{Appendix \thesection\protect\indent \parbox[t]{11.15cm}{#1}}
\addcontentsline{toc}{section}{Appendix \thesection\ \ \ #1}}

\def\bbl{{\bf{\ell}}}
\font\mybb=msbm10 at 11pt

\def\bb#1{\hbox{\mybb#1}}

\def\bZ {\bb{Z}}
\def\bR {\bb{R}}

\def\ua{\underline{a}}
\def\ub{\underline{b}}
\def\uc{\underline{c}}

\def\ur{\underline{r}}
\def\us{\underline{s}}
\def\ut{\underline{t}}

\def\cD{{\cal D}}
\def\tell{\hat{\ell}}
\def\tr{\hat{r}}

\newcommand{\bea}{\begin{eqnarray}}
\newcommand{\eea}{\end{eqnarray}}

\usepackage[usenames,dvipsnames,svgnames,table]{xcolor}	
\usepackage[normalem]{ulem}				
\usepackage{microtype}
\usepackage[colorlinks, citecolor=blue, linkcolor=blue, urlcolor=Maroon, filecolor=Maroon, linktocpage=true]{hyperref} 

\usepackage{chngcntr}
\counterwithout{equation}{section}

\begin{document}

\begin{center}
\vspace*{-1.0cm}
\begin{flushright}
\end{flushright}

\vspace{2.0cm} {\Large \bf Symmetries, a systematic construction of invariant fields and   AdS backgrounds } \\[.2cm]

\vskip 2cm
 G.  Papadopoulos
\\
\vskip .6cm


\begin{small}
\textit{Department of Mathematics
\\
King's College London
\\
Strand
\\
 London WC2R 2LS, UK}\\
\texttt{george.papadopoulos@kcl.ac.uk}
\end{small}
\\*[.6cm]

\end{center}

\vskip 2.5 cm

\begin{abstract}
\noindent
We   give a systematic local description of invariant metrics and other invariant fields on a spacetime under the action of a (non-abelian) group. This includes the invariant fields
 in a neighbourhood of a principal and a special orbit.  The construction is illustrated with  examples.  We also apply the formalism to give the R-symmetry invariant metrics of some AdS backgrounds
and comment  on applications to Kaluza-Klein theory.

\end{abstract}

\newpage

\renewcommand{\thefootnote}{\arabic{footnote}}


\section{Introduction}

It is well known that most explicit solutions of  Einstein equations    are invariant under  some group acting on the spacetime. These include black holes and cosmological solutions as well as large classes of solutions that have applications in string theory and AdS/CFT correspondence.  In this article, we shall propose a systematic way to construct local invariant metrics  given the action of  a (non-abelian) group  on a spacetime. Similar results are obtained for other invariant fields. The most general local expressions for the invariant metric and forms under the action of a group $G$ are given in (\ref{invmetr}) and (\ref{informsx1}), respectively. To our knowledge, these expressions include all ansatzes used in the literature to describe such  invariant fields  under the action of a group.

The construction presented below applies  generally to all problems that require a description of invariant fields under some group action. Nevertheless the initial  motivation for this work has been the classification of AdS backgrounds in 10- and 11-dimensional supergravity theories which have applications in AdS/CFT correspondence, for a review see \cite{maldacena}.   It is known that
 AdS$_n$ backgrounds admit a $SO(n-1,2)\times G$ group of symmetries  that leave all  fields invariant, where  the group\footnote{In supersymmetric backgrounds, $G$ includes the R-symmetry group.} $G$ acts on the internal space of the background.  The groups $G$ are compact  and  their action on the
internal space of an AdS solution is non-linear. The main difficulty to give an expression for the invariant fields under some group action is that it requires a model of  how a group  acts on a manifold. This is resolved with the application of slice and principal orbit theorems which we shall describe below.

There is much progress in the classifications of supergravity  AdS solutions. In particular the maximally supersymmetric AdS solutions of 10- and 11-dimensional supergravities have been classified in \cite{maxsusy}. Moreover  those that preserve strictly more than 16 supersymmetries have been classified in \cite{bgp, hlp1, hlp2, ggp} using either global methods or the homogeneity theorem of \cite{homogen} and the classification of all homogenous spaces up to dimension 9, see e.g. \cite{klausthesis}. Some progress has also been made towards the classification of AdS solutions that preserve 16 supersymmetries. In particular, there is a partial classification\footnote{In \cite{Mads7} and \cite{IIBads7}, it is assumed that the Killing spinors factorize as a suitable product of Killing spinors on AdS and Killing spinors on the internal manifold.} of such  AdS$_7$ backgrounds \cite{Mads7, IIBads7, bgp2}. Moreover,  there are no smooth AdS$_6$ solutions with compact without boundary internal space  in 11-dimensional and (massive) IIA supergravities\footnote{Non-existence theorems for AdS$_6$ backgrounds have also been proved in \cite{tomasielloads6, passias} under some additional assumptions.}\cite{ads6}. Furthermore if IIB AdS$_6$ backgrounds exist, the R-symmetry group, which has Lie algebra $\mathfrak{so}(3)$, must  have codimension 2 principal orbits as well as a non-empty set of special orbits and both IIB scalars, axion and dilaton, must  be non-constant functions on the internal space.  However, there are several AdS$_6$ solutions  with non-compact internal space and/or with singularities, see e.g. \cite{andy}-\cite{passias}.  For the remaining AdS backgrounds systematic results are sparse although many explicit solutions are known with widespread applications, see e.g \cite{rev} for  review and references within.

Further progress on the classification of supersymmetric AdS backgrounds, especially those that preserve less than half of supersymmetry, depends on  understanding how
the R-symmetry groups act on  the internal spaces of such a solution. The Lie algebras of all R-symmetry groups of warped AdS backgrounds under some mild assumptions have been found in \cite{superalg}. This together with the classification of all homogenous spaces up to dimension 9, see e.g \cite{klausthesis}, allow for the identification of all orbit types of the R-symmetry group in the internal space of the backgrounds up to a possible discrete identification.  We shall use these data to give a systematic local construction of   invariant metrics on the internal space of supersymmetric  AdS backgrounds.

This paper is organized as follows. In section two, the slice and principal orbit theorems are described. In section three, our main result is given after a detailed description of the invariant geometry of $G/H$ spaces which is required for the proof. In section four, some examples of group actions are given with orbits
$S^2$ and $S^3$ which illustrate the use of slice theorem. Then some applications to AdS$_n$ backgrounds, for $n=6,5,4$ are presented. In section 5, we give our conclusions and comment on applications to Kaluza-Klein theory.

\section{Preliminaries}

\subsection{Slice and principal orbit theorems}

The properties of a group action on a manifold  have been extensively investigated \cite{gleason, koszul, mont, palais}, see  \cite{brendt} for a recent review and also \cite{gibbons} for other applications. The following two main results  will be used here.

 \begin{enumerate}

 \item The  {\it slice theorem } or  tubular equivariant neighbourhood  theorem   states under some compactness assumptions\footnote{For example  assume that   $G$ and $M$ are compact which is sufficient for the applications to AdS backgrounds.} that the neighbourhood of an orbit $N$ of a group $G$ acting smoothly on a manifold $M$ is equivariantly diffeomorphic to  an invariant vector bundle $E$ over the orbit.  $E$ is identified  with  the normal bundle of $N$ in $M$.  This in particular means that the group action on a manifold can  locally be modelled as a lift of the action of $G$ on $N$ to $E$.

     \item
     The {\it principal orbit theorem} states that the union of all principal orbits\footnote{Principal orbits are those that have the smallest isotropy group in $G$ up to a conjugation.}, i.e. those of maximal dimension,  is a dense set of the manifold.  In addition away from some special orbits, the manifold is a bundle with fibre the typical principal orbit $N$ and base space $B$.

\end{enumerate}

The slice theorem provides a model of how a group $G$ acts on a neighbourhood of an orbit $N$ in $M$.  As $G$ acts transitively on $N$, $N$ can be identified with a homogeneous space, $N=G/H$. The slice theorem then states that on a neighbourhood of a $G/H$ orbit one can adapt  as coordinates those of $E$, and that on the fibre coordinates of $E$, $G$ acts with a rotation.  This gives a very concrete description of the vector fields generated
by the group action in the neighbourhood of $G/H$.   In turn this allows for the identification  of the invariant metrics and other invariant fields in the vicinity  of $G/H$ in $M$.

To give a very brief sketch of the idea behind the slice theorem, suppose that $N$ is a point $p$ and $M$ is equipped with a $G$-invariant metric $ds^2$. First consider  the exponential map $\exp: V\subset T_pM \rightarrow W\subset M$, that is   a diffeomorphism of the open sets  $V$ and $W$, $p\in W$, constructed using the geodesics of $ds^2$ parameterized with the affine length. As the metric $ds^2$ is invariant under the action $\alpha_g$ of $G$,   $\alpha_g: W\rightarrow W$, $\alpha_g$ will map geodesics of $ds^2$ to the geodesics of $ds^2$ for every  $g\in G$. So for any two points $q, z\in W$ with $\alpha_g(q)=z$, one has $\alpha_g(\exp v_q)=\exp v_z$, where $v_q, v_z\in V $ with $\exp v_q=q$ and $\exp v_z=z$. As the distance between $p$ and $q$, and $p$ and $z$ is the same as a consequence of the invariance of metric under $G$, and the geodesics are straight lines in the coordinates of $V$, $v_q$ and $v_z$ have the same length. As a result the induced action $\exp^{-1}\circ \alpha_g \circ \exp $ on $V\subset T_pM$ is an orthogonal transformation.  Thus the action of $G$ on a neighbourhood $V$ of the fixed point $p$ is modelled with rotations.

 One of the consequences of the  principal orbit theorem is that the normal bundle of a principal orbit $G/H$ is topologically trivial.  This is because it can be identified  with the pull back
of the tangent bundle of the base space $B$ restricted on the orbit $G/H$.  Another consequence is that one should only consider as principal orbits $G/H$ spaces
for which $G$ acts (almost) effectively on $G/H$.  Otherwise, $G$ will not act (almost) effectively on $M$.  However, $G$ may not act effectively on special orbits, e.g. $G$ does not act effectively on fixed points.

\subsection{A Frobenius approach}

Before, we proceed with the use of the above two theorems to describe the invariant geometry of a manifold, let us consider an (almost) effective action of a group
 $G$ on a  $M$ which  generates the vector fields $\xi_{\ur}$, $\ur=1, \dots, \mathrm{dim}\,\mathfrak{g}$, where $\mathfrak{g}$ is the Lie algebra of $G$. The Lie bracket algebra of these vector fields closes as
\bea
[\xi_{\ur}, \xi_{\us}]=f_{\ur\us}{}^{\ut}\, \xi_{\ut}~,
\label{comlie}
\eea
where $f$ are the structure constants of $\mathfrak{g}$.   The task is to write down the most general form of a metric on $M$ which is invariant under the action of $\xi$. For this let us explore first the  Frobenius theorem. In particular assume that there is an open subspace $U\subseteq M$ such that $\xi$'s span a subbundle $L$ of $TU$ of rank $k$. The Frobenius theorem states that the $\xi$'s define a regular foliation. This means that there is a submanifold $N$ of $U$, called the leaf of the foliation, such that  $TN=L\vert_N$. Moreover $U$ admits an atlas with coordinates $x^M=(w^I, z^m)$ and  patching conditions $w^I_\alpha=w^I_{\alpha\beta}( w^J_\beta, z^m_\beta), z^m_\alpha=z^m_{\alpha\beta}(z^n_\beta)$ such that
\bea
\xi_{\ur}=\xi^I_{\ur}(w^J, z^n)\, \partial_I~,~~~I=1,\dots, k~.
\label{xiv}
\eea
Note that the components of the vector fields may depend on both the coordinates $w$ of the leaf $N$ and  $z$ of the base space $B$ of the bundle $L$.

Using these coordinates the most general metric on $U$ can be written as
\bea
ds^2=g_{IJ}(dw^I+\Gamma^I_m\, dz^m) (dw^J+\Gamma_n^J\, dz^n)+ \gamma_{mn}\, dz^m dz^n~,
\label{metr}
\eea
where all components of the metric $\gamma$, $\Gamma$ and $g$ depend on both $w$ and $z$ coordinates.  The metric retains its form under the patching conditions provided that
\bea
&&\gamma_{\alpha}(x^M_\alpha)_{ mn}={\partial z_\beta^p\over \partial z_\alpha^m}{\partial z_\beta^q\over \partial z_\alpha^n} \gamma_\beta(x^M_\beta)_{pq}
~,~~~
g_\alpha(x^M_\alpha)_{IJ}={\partial w_\beta^K\over \partial w_\alpha^I}{\partial w_\beta^L\over \partial w_\alpha^J} g_\beta(x^M_\beta)_{KL}~,
\cr
&&\Gamma_\alpha(x^M_\alpha)^I{}_m={\partial z_\beta^n\over \partial z_\alpha^m}{\partial w_\alpha^I\over \partial w_\beta^J}\Gamma_\beta(x^M_\beta)^J{}_n- {\partial w_\alpha^I\over \partial z_\beta^n}{\partial z_\beta^n\over \partial z_\alpha^m}~.
\eea
Clearly $g$ restricted on $N$ is a metric on $N$ and $\gamma$ is a fibre metric on $\pi^*TB$, where $\pi$ is the projection $\pi: U\rightarrow B$. Furthermore $\Gamma$ is a non-linear connection \cite{vilms}  and  gives a splitting $j:\pi^*TB\rightarrow TU$ of the sequence
\bea
0\rightarrow TN\rightarrow TU\rightarrow \pi^*TB\rightarrow 0~.
\eea
In particular the horizontal vector fields which are identified with the sections of $j\pi^*TB$ in the decomposition $TU=TN\oplus j\pi^*TB$ are
$j(X)=X^m\partial_m-X^m \Gamma_m^I\partial_I$, where $X^m\partial_m$ is a section of $TB$.

Imposing the invariance of the metric (\ref{metr}) under the action of the vector fields (\ref{xiv}), one finds that
\bea
&&\xi_{\ur}^I \partial_I g_{JK}+\partial_J\xi_{\ur}^I g_{IK}+\partial_K\xi_{\ur}^I g_{JI}=0~,~~~\xi_{\ur}^I \partial_I \gamma_{mn}+\partial_m\xi_{\ur}^I g_{IJ} \Gamma^J_n+\partial_n\xi_{\ur}^I g_{IJ} \Gamma^J_m=0~,
\cr
&&
\xi_{\ur}^J \partial_J \Gamma^I_m- \partial_J \xi_{\ur}^I \Gamma^J_m+ \partial_m \xi_{\ur}^I=0~.
\label{kcon}
\eea
The above conditions on the components of the metric can be simplified dramatically if the vectors fields $\xi$ can be arranged to be independent of $z^m$ coordinates. In particular $\gamma$ would have been independent of the $w$ coordinates and so it would have been the pull back of a metric on $B$ and the non-linear
connection $\Gamma$ would have been strictly invariant under the action of $G$ instead of being invariant up to a gauge transformation as above.  However in general, the isometry conditions (\ref{kcon}) do not have an explicit solution unless some additional assumptions are made on the
vector fields $\xi$.  As a result, it is not apparent how to explicitly express  an invariant metric under a group action on a manifold in this approach.

 Before we proceed to resolve this puzzle using the slice and principal orbit theorems, notice that locally the metric (\ref{metr}) can be rewritten as
\bea
ds^2= \tilde g_{IJ} dw^I dw^J+ \tilde \gamma_{mn} (dz^m+\tilde\Gamma^m_I dw^I)(dz^n+\tilde\Gamma^n_J dw^J)~,
\label{metrf2}
\eea
where
\bea
&&\tilde g_{IJ}+\tilde \gamma_{mn} \tilde\Gamma^m_I\tilde\Gamma^n_J=g_{IJ}~,~~~\tilde \gamma_{mn} \tilde \Gamma^m_J=g_{IJ} \Gamma^I_n~,~~~\tilde \gamma_{mn}=\gamma_{mn}+g_{IJ} \Gamma^I_m\Gamma^J_n~.
\eea
In this form, the metric is locally adapted again to a fibration but now with fibre $B$ and base space $N$. Furthermore the new fibre twists over $N$ with connection $\tilde \Gamma$. It turns out that this is the approach taken by the slice theorem.

Concluding this section it  is worth emphasizing  that the coordinates adapted on the manifold via the use of the slice theorem and those adapted via the use  of the Frobenius theorem are different and so in what follows they will be denoted differently.
As we have seen in the context of the Frobenius theorem, we have set $x^M=(w^I, z^m)$, where $w$ are the coordinates of the leaf and $z$ are the coordinates of the base space. While in the context of the slice theorem,  we shall denote the linear fibre coordinates   of the vector bundle $E$ with $y^a$.  The metric and other invariant tensors on $M$ depend on the coordinates of $G/H$  via that of left-invariant frame $\ell$ and the canonical connection $\Omega$ defined in (\ref{MC}).

\section{Systematic construction of invariant fields}

\subsection{Lifting group actions}

 To explore further the consequences of the slice theorem, let us investigate the  invariant vector bundles $E$ over $G/H$.  These are bundles for which the left action $a_g$ of $G$, $g\in G$, on $G/H$   can be lifted to the bundle space $E$. Before we investigate the lifting of group actions to vector bundles, let us first consider the lifting of group actions to principal bundles. A principal bundle $P(K)$ on $G/H$ with fibre $K$  admits a lifting $a_g^\uparrow: P(K)\rightarrow P(K)$ of the left action $a_g$  on $G/H$ iff there is a group action $a_g^\uparrow$ of $G$ on the bundle space $P(K)$ such that $\pi\circ a_g^\uparrow=a_g\circ \pi$ and $a_g^\uparrow(pk)=a_g^\uparrow(p) k$, where $\pi: P(K)\rightarrow G/H$ is the projection, $p\in P(K)$ and $k\in K$.    Denote with $P(H)$ the ``master'' principal bundle which arises from the right action of the subgroup $H$ on $G$, $H\rightarrow G\rightarrow G/H$. Clearly the bundle space of $P(H)$ is $G$.  One can demonstrate the following \cite{yoshida, kobayashi,  isham}.
\begin{enumerate}

\item All principal bundles $P(K)$ that admit a lifting $a_g^\uparrow$ of the left action $a_g$  of $G$ on $G/H$ are associated bundles\footnote{We also denote $G\times_\alpha K$ with $G\times_H K$, and $P(K)\times_D V$ below with   $P(K)\times_K V$ at convenience.} $P(K)=G\times_\alpha K$ of the master principal bundle
 $P(H)$, where $\alpha: H\rightarrow K$ is a group homomorphism given by $a_h^\uparrow(p_0)=p_0 \alpha(h)$, $h\in H$ and $p_0$ a fixed point in $P(K)$ with $\pi(p_0)=eH$. Note that
 $ K\rightarrow G\times_\alpha K\rightarrow G/H$
 and  if $(g, k)_\alpha\in G\times_\alpha K$, then $(g, k)_\alpha=\big(gh, \alpha(h^{-1})k\big)_\alpha$ for any $h\in H$.  Moreover the lifted action of $G$ on $G\times_\alpha K$ is  $a_{g'}^\uparrow(g, k)_\alpha=(g'g,k)_\alpha$.

    \item The lifting ${a}^\uparrow{}_{g}$ of $a_g$ to an associated bundle $E_D=P(K)\times_D V$ of $P(K)$, where $D$ is a representation of $K$ on the vector space $V$, is
    ${a}^\uparrow{}_{g}(p, v)_D=({a}^\uparrow{}_{g}(p), v)_D$, where we have used ${a}^\uparrow{}_{g}$ to denote  the lift of $a_g$ on both $G\times_\alpha K$ and $E_D$.  Note that if $(p, v)_D\in E_D$, then $(pk, D(k^{-1})v)_D = (p, v)_D$ for any  $k\in K$.
\end{enumerate}

 Typically  there are several inequivalent lifts $a_g^\uparrow$ of $a_g$ on $G/H$ to a principal bundle  $P(K)$ over $G/H$ \cite{yoshida, isham}.  For every such lift $a_g^\uparrow$, there is a group homomorphism $\alpha$ such that   ${a}^\uparrow{}_{g'} (g, k)_\alpha=(g'g, k)_\alpha$. In particular viewing $E_D$ as an associated bundle of $P(H)$, $E_D=G\times_{\alpha\circ D} V$, the vector fields generated on $E_D$ are
 \bea
 \xi_{\ur}=q_* R_{\ur}~,
 \eea
 where $q: G\times V\rightarrow E_D$ is the standard projection, $q_*$ is the push-forward map and $R_{\ur}$ are the right-invariant vector fields on $G$.

The identification of a neighbourhood $U$ of a $G/H$ orbit in a manifold $M$ with an invariant bundle $E_D$ over $G/H$ and the identification of the action of $G$ on $U$ with the ${a}^\uparrow{}_{g}(p, v)_D=({a}^\uparrow{}_{g}(p), v)_D$  of $G$ on $E_D$ has several consequences. One is that the vector fields $\xi$ generated by the action of $G$ on $U$ can be explicitly written down
in terms of the right invariant vector fields on $G/H$ and those generated on the fibres of $E_D$ by the representation $\alpha\circ D$ of $H$.  The simplicity of the group action of $G$ in these coordinates allows for the systematic construction of the invariant metrics on $U\subset M$ utilizing the invariant geometry on $G/H$ and $E_D$.

\subsection{Invariant geometry of homogeneous spaces}\label{secx1}

Having identified $E_D$ as the neighbourhood of an orbit of $G$ in $M$, it remains to construct the most general invariant metric and other invariant fields  on $E_D$.  As in many applications of interest, like in AdS/CFT, the Lie algebra of isometries is known instead of the group. Because of this, we shall focus on the construction of the invariant fields under the action of a Lie algebra. The construction presented below can be adapted to include invariance of the fields  under  groups too but we shall not explore this further here, see e.g.  \cite{kobayashi}.   To proceed take $G/H$ to be reductive\footnote{Some of our constructions can be extended to the non-reductive case. In any case, if $G$ and $H$ are compact, then it can  always be arranged such that $G/H$ is reductive.} so that the commutators of $\mathfrak{g}=\mathfrak{h}\oplus \mathfrak{m}$ are
\bea
[\mathfrak{h},\mathfrak{h}]\subseteq \mathfrak{h}~,~~~[\mathfrak{h},\mathfrak{m}]\subseteq \mathfrak{m}~,~~~[\mathfrak{m},\mathfrak{m}]\subseteq \mathfrak{h}\oplus\mathfrak{m}~,
\eea
where $\mathfrak{h}$ is the Lie algebra of $H$.

Denote   the generators of  $\mathfrak{h}$ with $h_\alpha$, $\alpha=1,2,..., \dim{\mathfrak{h}}$  and a basis in $\mathfrak{m}$ as  $m_A$, $A=1,..., \dim{\mathfrak{g}}-\dim{\mathfrak{h}}$. In this basis,  the brackets of the Lie algebra $\mathfrak{g}$ take the following form

\bea\label{commutation}
&&[h_\alpha, h_\beta] = f_{\alpha\beta}{}^\gamma \, h_\gamma~,~~~
[h_\alpha, m_A] = f_{\alpha A}{}^B \, m_B~,
\cr
&&[m_A,m_B] = f_{AB}{}^C \, m_C + f_{AB}{}^\alpha \, h_\alpha~.
\eea
If $f_{AB}{}^C=0$, that is $[\mathfrak{m},\mathfrak{m}] \subset \mathfrak{h}$, the space is symmetric.

Let $\theta$ be the Maurer-Cartan form on $G$. Thus  $\theta(X_A)=A$ for the  left-invariant vector field $X_A$  generated by the right action of $A\in \mathfrak{g}$ on $G$.  One can define a canonical connection and a frame on $G/H$ by decomposing  $\theta$ along  $\mathfrak{h}$ and  $\mathfrak{m}$ on $G$ and then pulling back the resulting expression on $U\subset G/H$ with the local section $s: U\subset G/H\rightarrow G$.  In particular, one has
\begin{align}
s^*\theta=s^{-1} ds=  \bbl^A \, m_A + \Omega^\alpha \, h_\alpha~,
\label{MC}
\end{align}
where $\bbl^A$ is  a local left-invariant frame and $\Omega^\alpha$ the canonical left-invariant connection.  Note that $\Omega=s^*\theta\vert_{\mathfrak{h}}$, where $\theta\vert_{\mathfrak{h}}$ is the canonical principal bundle connection on $H\rightarrow G\rightarrow G/H$, see e.g. \cite{kobayashi}.  The curvature and torsion of the canonical connection are
\bea\label{dei}
&&R^\alpha \equiv d\Omega^\alpha+\frac12 f_{\beta\gamma}{}^\alpha \Omega^\beta\wedge \Omega^\gamma=-\frac12 f_{BC}{}^\alpha \bbl^B\wedge \bbl^C~,
\cr
&&T^A\equiv d\bbl^A+f_{\beta C}{}^A \Omega^\beta\wedge \bbl^C=-\frac12 f_{BC}{}^A \bbl^B\wedge \bbl^C~,
\eea
respectively, where the equalities follow after  taking the exterior derivative of (\ref{MC}) and  using (\ref{commutation}).  If $G/H$ is a symmetric space, then
the torsion vanishes.

A left-invariant   p-form $\omega$ on $ G/H$ can be written as
\begin{align}
\omega = \frac{1}{p!} \, \omega_{A_1 ... A_p} \, \bbl^{A_1} \wedge ... \wedge \bbl^{A_p}~,
\end{align}
where the components   $\omega_{A_1...A_p}$ are constant and satisfy
\begin{align}\label{hinvariance}
f_{\alpha[A_1}{}^B \, \omega_{A_2...A_p]B} =0~.
\end{align}
The latter condition is required for invariance under the right action of $H$ on $G$. All left-invariant forms are  parallel with respect to the canonical connection.

It remains to describe the metrics of $G/H$ which are  left-invariant.  These are written as
\bea
ds^2=g_{AB}\, \bbl^A \bbl^B~,
\eea
where the components $g_{AB}$ are constant and satisfy
\begin{align}\label{ginvar1}
f_{\alpha A}{}^C \, g_{BC} + f_{\alpha B}{}^C \, g_{AC} = 0~.
\end{align}
For symmetric spaces, the canonical connection coincides with the Levi-Civita connection of invariant metrics.

To present the invariant metrics on $E_D$, one also needs to describe the invariant connections on $P(K)$. For this  let us denote with  $\rho$ the Lie algebra homomorphism $\rho: \mathfrak{h}\rightarrow \mathfrak{k}$ induced from the
Lie group homomorphism $\alpha: H\rightarrow K$ which characterizes the principal bundle $P(K)=G\times_\alpha K$.  Next consider
the linear map $\Lambda: \mathfrak{m}\rightarrow \mathfrak{k}$ such that
\bea
f_{\alpha A}{}^B \Lambda_B^{\ua}=\rho_\alpha^{\uc} f_{\uc\ub}{}^{\ua} \Lambda^{\ub}_A~,
\label{ginvar2}
\eea
where $f_{\uc\ub}{}^{\ua}$ are the structure constants of the Lie algebra of $K$, $\mathfrak{k}$, in a basis $t_{\ua}$.  Then the most general linear invariant connection
is
\bea
\Sigma^{\ua}=\rho^{\ua}_\alpha\, \Omega^\alpha+ \Lambda_A^{\ua}\, \ell^A~,
\eea
where  $\rho^{\ua}_\alpha\,\Omega^\alpha$ is the canonical connection of $P(K)=G\times_\alpha K$. For a proof of this and more details,  see e.g \cite{kobayashi}.

The curvature $F$ of $\Sigma$ is
\bea
F^{\ua}\equiv d\Sigma^{\ua}+{1\over2}f_{\ub\uc}{}^{\ua}\, \Sigma^{\ub}\wedge \Sigma^{\uc}=-\frac12\, \rho^{\ua}_\alpha\, f_{BC}{}^\alpha \ell^B\wedge \ell^C
+\left(f_{\ub\uc}{}^{\ua} \Lambda^{\ub}_B \Lambda^{\uc}_C  -\Lambda^{\ua}_A f_{BC}{}^A \right)\ell^B\wedge \ell^C~.
\eea
 The induced connection on $E_D=P(K)\times_D V$  is as usual
\bea
\Sigma^a{}_b=\Sigma^{\ua}\,  \cD_{\ua}{}^a{}_b~,
\eea
where $\cD$ is the representation of $\mathfrak{k}$ on $V$ induced by $D$.

To construct the most general class of metrics invariant under $G$ on $M$, let us consider
\bea
\Pi^a=\Pi^a_A \ell^A~,
\eea
where $\Pi^a_A$ are constants,
and impose the condition
\bea
f_{\alpha A}{}^B \Pi_B^a= \rho^{\ua}_\alpha\, \cD_{\ua}{}^a{}_b\Pi_A^b~.
\label{ginvar4}
\eea
This condition is required for $\Pi$ to transform covariantly under the right $H$ transformations.

Clearly a fibre metric on $P(K)\times_D V$ will be invariant under the action of $G$ iff
\bea
\gamma_{c b}\, \rho^{\ua}_\alpha\, \cD_{\ua}{}^c{}_a+\gamma_{c a}\, \rho^{\ua}_\alpha\, \cD_{\ua}{}^c{}_b=0~.
\label{ginvar3}
\eea
For completeness, the invariant forms with values in the tensor product of bundle $\otimes^q (G\times_K V)$ satisfy the condition
\bea
\label{pqinv}
&&p (-1)^{p-1} f_{\alpha[A_1}{}^B \, \omega_{A_2...A_p]Ba_1a_2\dots a_q}+ \rho^{\ua}_\alpha \cD_{\ua}{}^b{}_{a_1}\omega_{A_1\dots A_p b a_2\dots a_q}
\cr
&&
+ \rho^{\ua}_\alpha \cD_{\ua}{}^b{}_{a_2}\omega_{A_1\dots A_p  a_1b\dots a_q}+\dots + \rho^{\ua}_\alpha \cD_{\ua}{}^b{}_{a_q}\omega_{A_1\dots A_p a_1 a_2\dots b} =0~.
\eea
All the above formulae can be easily adapted to the special case where $K=H$ and $\rho=1_{\mathfrak{h}}$.

\subsection{Main result: Invariant fields on the spacetime}

 In the context of homogeneous spaces the metric $g_{AB}$, connection $\Lambda^{\alpha}_A$, $\Pi^a_A$ and fibre metric
$\gamma_{ab}$ are all constants.  To construct a    metric on $E_D=G\times_H V$ invariant under $G$ and so suitable to model  an  invariant metric
on $M$ in the neighbourhood of an orbit, we allow  $g_{AB}$,  $\Lambda^{\alpha}_A$, $\Pi^a_A$ and
$\gamma_{ab}$ to depend on $y$, i.e. become functions of the fibre $V$.   With this understanding, the metric
\bea
ds^2=g_{AB} \ell^A \ell^B+ \gamma_{ab} (dy^a+  \Sigma^{\ua}\, \cD_{\ua}{}^a{}_c \,y^c+\Pi^a_A\, \ell^A)(dy^b+  \Sigma^{\ub}\, \cD_{\ub}{}^b{}_d\, y^d+\Pi^b_B \, \ell^B)~,
\label{invmetr}
\eea
on $G\times_HV$  is invariant under the action of $G$ provided that
\bea
&&{\cal L}_{\cD_\alpha} \Lambda^\beta_A +f_{\alpha A}{}^B \Lambda_B^{\beta}- f_{\alpha\gamma}{}^{\beta} \Lambda^\gamma_A=0~,
\cr
&&{\cal L}_{\cD_\alpha}\Pi_B^a+ f_{\alpha A}{}^B \Pi_B^a-  \cD_{\alpha}{}^a{}_b\Pi_A^b=0~,
\cr
&&{\cal L}_{\cD_\alpha}\gamma_{ab}+\gamma_{c b}\, \cD_{\alpha}{}^c{}_a+\gamma_{c a}\,  \cD_{\alpha}{}^c{}_b=0~,
\cr
&& {\cal L}_{\cD_\alpha}g_{AB}+f_{\alpha A}{}^C \, g_{BC} + f_{\alpha B}{}^C \, g_{AC} =0~,
\label{ginvar4xxx}
\eea
 where $\cD_\alpha{}^a{}_b= \rho^{\ua}_\alpha \cD_{\ua}{}^a{}_b$ and ${\cal L}_{\cD_\alpha}$ denotes the Lie derivative with respect to the vector fields ${\cal D}_\alpha=\rho^{\ua}_\alpha \cD_{\ua}{}^a{}_b y^b\partial_a$ generated by the action $\cD$ of $\mathfrak{h}$  on $V$.

The conditions (\ref{ginvar4xxx}) are the analogues to those in (\ref{kcon}) but there is a difference. Here the vector field ${\cD_\alpha}$ appearing in the Lie derivative ${\cal L}_{\cD_\alpha}$ is that of a rotation.  Because of this, it is more straightforward to solve (\ref{ginvar4xxx}) instead of (\ref{kcon}).

A similar construction works for other fields.  In particular an invariant form under the group action of $G$ in the neighbourhood of an orbit $G/H$  is
\bea
\omega={1\over p!\, q!}  \omega_{A_1\dots A_p a_1 \dots a_q}\, \ell^{A_1}\wedge\dots\wedge \ell^{A_p}\wedge e^{a_1}\wedge\dots \wedge e^{a_q}~,
\label{informsx1}
\eea
provided that
\bea
\label{pqinvy}
&& {\cal L}_{\cD_\alpha}  \omega_{A_1\dots A_p a_1 \dots a_q}+  p (-1)^{p-1} f_{\alpha[A_1}{}^B \, \omega_{A_2...A_p]Ba_1a_2\dots a_q}+  \cD_{\alpha}{}^b{}_{a_1}\omega_{A_1\dots A_p b a_2\dots a_q}
\cr
&&
+  \cD_{\alpha}{}^b{}_{a_2}\omega_{A_1\dots A_p  a_1b\dots a_q}+\dots +  \cD_{\alpha}{}^b{}_{a_q}\omega_{A_1\dots A_p a_1 a_2\dots b} =0~,
\eea
where $e^a=dy^a+  \Sigma^{\ua}\, \cD_{\ua}{}^a{}_c \,y^c+\Pi^a_A\, \ell^A$.

  A special solution to   (\ref{ginvar4xxx}) and (\ref{pqinvy}) can be constructed as follows. In the context of homogeneous spaces, the conditions (\ref{ginvar1}), (\ref{ginvar2}), (\ref{ginvar4}), (\ref{ginvar3}) and (\ref{pqinv}) determine $\Lambda$, $\gamma$, $\Pi$,  $g$ and $\omega$ up to some constants. Now if one {\it replaces those constants}  with {\it invariant functions} under the representation $D$ of $H$ on $V$, the conditions
 (\ref{ginvar4xxx}) and (\ref{pqinvy})  will be automatically satisfied.  This is an assumption made in all examples demonstrated below.

Although the metric proposed in (\ref{invmetr}) is constructed under some smoothness assumptions for  both the group action and the associated underlying manifold, it is a good starting point to investigate solutions to the Einstein equations even in the case that these assumptions are violated.  In fact it is expected that many solutions to the field equations that can be constructed using (\ref{invmetr}) will be singular. In the presence of singularities, the slice and  principal orbit theorems may not apply everywhere but locally (\ref{invmetr}) still can be used to find solutions.  Similarly, one can remove some of the compactness assumptions necessary for the validity of the slice and principal orbit theorems and still use (\ref{invmetr}) to identify some invariant metrics.

The metric  (\ref{invmetr}) describes the invariant geometry of the spacetime in the neighbourhood of an orbit $N$ independent on whether the orbit is principal or special\footnote{For special orbits, one should consider all homogeneous spaces $G/H=N$  including those that
$G$ does not act effectively on $N$.}. $N$ is identified with the zero section of $E_D$.   For principal orbits, we have already mentioned that as a consequence of the principal orbit theorem $E_D$ must be a topologically trivial bundle over $G/H$.
This is not a sufficient condition. In particular one has to demonstrate that the orbits of $G$ on $E_D$ away from the zero section are still $N$. This can be proven directly by identifying the isotropy group of the orbits of $G$ in a neighbourhood of the zero section in $E_D$   and compare it with the isotropy group of $G$ acting on $N$. If the isotropy subgroups in $G$ are isomorphic up to a conjugation, then this will prove that the orbit $N$ is principal. One can also compute the codimension of the orbits of $G$ in a neighbourhood of the zero section of $E_D$  and compare it with that of $N$ in $M$. For this test, it is equivalent to require that the equations
\bea
\xi_{\ur} f=q_* R_{\ur} f=0~,
\label{indsol}
\eea
have as many independent solutions as the codimension of $N$ in $M$, where $f$ is a function defined on a patch of $E_D$ and $\xi_{\ur}$ are the vector fields generated by the action of $G$ on $E_D$.
To see this observe that as a consequence of the Frobenius theorem, the vector field $\xi_{\ur}$ can be written as in (\ref{xiv}).  The independent solutions $f$ of (\ref{indsol}) are the coordinates $z$ of the spacetime, i.e. the coordinates of the base space $B$ of the foliation.

\section{Applications}

\subsection{Some examples}

In all examples that we shall investigate below as well as in all applications to AdS backgrounds, we shall assume that the ${\cal L}_{\cal D}$ terms in the conditions
(\ref{ginvar4xxx}) vanish. In the cases that $D$ is the trivial representation, this  follows automatically.  Otherwise, it is an additional assumption that we use to find solutions. Assuming this, the remaining conditions in  (\ref{ginvar4xxx}) are algebraic and can be solved.  As it has already been mentioned the end result is that components of the invariant fields will  depend on invariant functions of the coordinates $y$ of $V$ under  the action of the $D$ representation of $H$ on $V$.

\subsubsection{Invariant geometry on $SU(2)$ and $S^2$}

Before we proceed to apply the formalism developed so far to AdS backgrounds let us consider  examples mainly focused  on    $S^2$ and $S^3$ orbits.    To describe the most general $SU(2)$ invariant metric on a manifold with  $S^2=SU(2)/U(1)$ orbits, parameterize $SU(2)$ in terms of the Hopf coordinates as
\bea
g=\begin{pmatrix}e^{i\vartheta_1}\sin\eta& e^{i\vartheta_2} \cos\eta\\ -e^{-i\vartheta_2}\cos\eta&e^{-i\vartheta_1}\sin\eta\end{pmatrix}~,~~~g\in SU(2)~,
\eea
where $0\leq \eta\leq {\pi\over2}$ and $0\leq \vartheta_1, \vartheta_2\leq 2\pi$.
Choosing as a basis in $\mathfrak{su}(2)$ the anti-Hermitian matrices $\{t_1=i\sigma_1, t_2=i\sigma_2, t_3=i\sigma_3\}$,  where $\sigma_1, \sigma_2, \sigma_3$ are the Pauli matrices, the left invariant 1-forms are
\bea
&&\tell^3=-\cos(2\eta) d\rho-d\tau~,~~~\tell^2= -\cos(2\tau) d\eta-\sin(2\tau)\sin(2\eta) d\rho~,~~~
\cr
&&\tell^1=-\sin(2\tau) d\eta+ \cos(2\tau) \sin(2\eta) d\rho~,
\label{linvforms}
\eea
where $2\tau=\vartheta_2-\vartheta_1$ and $2\rho=\vartheta_1+\vartheta_2$.  In this basis, the left invariant vector fields are
\bea
&&L_3=-\partial_\tau~,~~~L_2=\sin(2\tau)\cot(2\eta)\partial_\tau-\cos(2\tau)\partial_\eta-{\sin(2\tau)\over\sin(2\eta)} \partial_\rho~,
\cr
&&L_1=-\cos(2\tau)\cot(2\eta)\partial_\tau-\sin(2\tau)\partial_\eta+{\cos(2\tau)\over\sin(2\eta)} \partial_\rho~.
\label{linvvf}
\eea
Note that $[L_r, L_s]=-\epsilon_{rs}{}^t L_t$, where $r,s,t=1,2,3$.
Similarly the right-invariant 1-forms are
\bea
&&\tr^3=d\rho+\cos(2\eta) d\tau~,~~~\tr^2=-\cos(2\rho) d\eta-\sin(2\rho) \sin(2\eta) d\tau~,~~~
\cr
&&
\tr^1=-\sin(2\rho) d\eta+\cos(2\rho) \sin(2\eta) d\tau~,
\label{rinvforms}
\eea
and the corresponding right-invariant vector fields are
\bea
&&R_3=\partial_\rho~,~~~R_2=\sin(2\rho) \cot(2\eta) \partial_\rho-\cos(2\rho) \partial_\eta-{\sin(2\rho)\over\sin(2\eta)}\partial_\tau~,
\cr
&&
R_1=-\cos(2\rho) \cot(2\eta) \partial_\rho-\sin(2\rho) \partial_\eta+{\cos(2\rho)\over\sin(2\eta)}\partial_\tau~,
\label{rightv}
\eea
where $[R_r, R_s]=\epsilon_{rs}{}^t R_t$.

To give the most general invariant metric on a manifold with  $S^2$ orbits, let us first investigate the invariant geometry on  $S^2=SU(2)/U(1)$.  For this assume without loss of generality that  the right $U(1)$ action on $SU(2)$   generates the left invariant vector field $L_3$.
If $\pi: SU(2)\rightarrow S^2=SU(2)/U(1)$ is the standard projection, then clearly $\pi_* L_3=\pi_*\partial_\tau=0$, where $\pi_*$ is the push forward map associated to   $\pi$. The left action of $SU(2)$ on $S^2$ generates the vector fields
\bea
&&\pi_* R_3=\partial_\rho~,~~~\pi_* R_2=\sin(2\rho) \cot(2\eta) \partial_\rho-\cos(2\rho) \partial_\eta~,~~~
\cr
&&
\pi_* R_1=-\cos(2\rho) \cot(2\eta) \partial_\rho-\sin(2\rho) \partial_\eta~,
\eea
where $\rho$ and $\eta$ are the coordinates of $S^2$.  As $[\pi_*X, \pi_*Y]=\pi_*[X,Y]$, the pushed forward vector fields $\pi_* R_s$, $s=1,2,3$,  satisfy the same
Lie algebra bracket relations as those of $R_s$, i.e. their Lie algebra is $\mathfrak{su}(2)$.

\subsubsection{Codimension two $S^2$ orbits}\label{s2orbits2}

As irreducible non-trivial real representations $D_n$,  $n\in \bZ-\{0\}$, of $U(1)$ are two dimensional, let us first assume that $S^2$ is a codimension 2 orbit.
Next consider the associated invariant vector bundle $E_n=SU(2)\times_{D_n}\bR^2$. The $U(1)$ action on $\bR^2$ via the  representation $D_n$ generates the vector field
\bea
{\cal D}_n=n(y^1{\partial\over \partial y^2}-y^2{\partial\over \partial y^1})~,
\eea
where $(y^1, y^2)$ are the standard coordinates of $\bR^2$. Note that in radial coordinates $y^1=r\cos\chi$ and $y^2=r\sin\chi$ on $\bR^2-\{0\}$
\bea
{\cal D}_n=n \partial_\chi~.
\eea
Let $p: SU(2)\times \bR^2\rightarrow E_n$ be the standard projection. It is clear that
$p_*(L_3-n\partial_\chi)=p_*(\partial_\tau+n \partial_\chi)=0$. The vector fields generated by the left action of $SU(2)$ on $E_n$, away from the zero section, are given by
\bea
&&\xi_3=p_*R_3=\partial_\rho~,~~~\xi_2=p_*R_2=\sin(2\rho) \cot(2\eta) \partial_\rho-\cos(2\rho) \partial_\eta+n {\sin(2\rho)\over\sin(2\eta)}\partial_\chi~,
\cr
&&
\xi_1=p_*R_1=-\cos(2\rho) \cot(2\eta) \partial_\rho-\sin(2\rho) \partial_\eta-n{\cos(2\rho)\over\sin(2\eta)}\partial_\chi~,
\eea
where $(\rho, \eta, \chi, r)$ are the coordinates of $SU(2)\times_{U(1)}\bR^2$ with $(\rho, \eta)$ the coordinates of the base space $S^2$.
Notice that $p_*R_1, p_*R_2, p_*R_3$  can have a non-trivial component along $\bR^2$.
If $q: SU(2)\times_{D_n}\bR^2\rightarrow S^2$ is the standard projection, then it is straightforward to observe that $q_*p_* R_s=\pi_* R_s$, $s=1,2,3$.   According to the slice theorem
$p_*R_1, p_*R_2, p_*R_3$ model the general action of $SU(2)$ at a neighbourhood of an $S^2$ orbit.

 To construct the most general invariant metric on $SU(2)\times_{U(1)}\bR^2$ first observe that the canonical connection $\Omega$ of $SU(2)/U(1)$ is $\Omega=s^*\tell^3=-\cos(2\eta) d\rho$, where the local section\footnote{This local section is chosen for convenience. Similar choices will be made in other examples below to write explicitly  the spacetime metric. However all choices of a local section  are equivalent as they are related by local gauge transformations.  Therefore the  choice of a particular section is not essential for the description of spacetime geometry.} $s: W\subset S^2\rightarrow SU(2)$ is chosen as $s(\eta, \rho)=(\eta, \rho, 0)$.
The condition on $\Lambda$ in (\ref{ginvar2}) implies that $\Lambda=0$. Furthermore if in addition $n\not=\pm1$, (\ref{ginvar4}) also implies that $\Pi=0$. So the only twisting of the fibre coordinates of the fibration is induced by the canonical connection. As $S^2$ is a symmetric space the invariant metric on $S^2$ is uniquely specified up to a constant $a$. Using these,  one finds that the invariant metric (\ref{invmetr}) can be written as
\bea
&&ds^2=a^2 \delta_{AB} \ell^A \ell^B+ b_1^2\, dr^2 +b_2^2\,  \big(d\chi- n \cos(2\eta) d\rho\big)^2 ~,
\label{metrs2}
\eea
where $\ell^A=s^*\tell^A$, $A=1,2$, and  $a, b_1, b_2$ depend only on the coordinate $r$ ($n\not=\pm1, 0$).  The fibre metric $\gamma$ decomposes as indicated
because of (\ref{ginvar3}).  For $n=\pm1$, there is an additional contribution from $\Pi$ in the metric (\ref{invmetr}) but this will not be explored here.

 For $n\not=0$, the vector bundles $SU(2)\times_{D_n}\bR^2$ are topologically non-trivial and so  do not model the neighbourhood of principal $S^2$ orbits of  codimension 2.
Alternatively, use (\ref{indsol}) and observe that the equations $\xi_rf=0$, $r=1,2,3$, for $n\not=0$,  have only one independent solution  instead of two required for codimension two orbits of $SU(2)$ in $SU(2)\times_{U(1)}\bR^2$.  The independent solution is $f=r$.  This again rules out $SU(2)\times_{U(1)}\bR^2$  as a neighbourhood for principal $S^2$ orbits. The metric (\ref{metrs2}) describes the geometry of a codimension one orbit of $SU(2)$ as approaches a codimension two special $S^2$ orbit.

On the other hand for $n=0$,
 $\xi_rf=0$, $r=1,2,3$, has two independent solutions $f=r,\chi$. The orbit  $S^2$ is principal.  Without loss of generality one can consider $S^2$ as codimension 1 principal orbit as $D_0$ is the trivial representation of $U(1)$ which is 1-dimensional. As $S^2=SU(2)/U(1)$ is a symmetric space, the invariant metric on $S^2$ is unique up to an overall scale.  The connection $\Sigma^{\ua} \cD_{\ua}$ vanishes as $\cD_n=0$ for $n=0$. Next focus on  the contribution that comes from  $\Pi$ in the metric (\ref{invmetr}).    $\mathfrak{h}=\mathfrak{u}(1)$ acts trivially on the fibre as $D_0=0$ but on the other hand acts with the fundamental 2-dimensional representation on $\mathfrak{m}$. As a consequence  (\ref{ginvar4}) implies that  $\Pi=0$. Thus the most general invariant metric that one can write down is
\bea
ds^2=a^2(y) \delta_{AB} \ell^A \ell^B+ b^2(y) dy^2~,
\label{ps2metr}
\eea
where $A,B=1,2$ and  $a^2, b^2$  are arbitrary functions of $y$. This is a warped  metric on $\bR\times S^2$, where $S^2$ is the round 2-sphere. This result can be easily generalized to orbits of codimension $\geq 1$, see also section \ref{secxx1}.

\subsubsection{Codimension four $S^2$ orbits}

Next consider codimension 4 principal $S^2$ orbits.  As any real representation of $U(1)$ decomposes to a direct sum of $D_n$ representations, $SU(2)\times_{U(1)}\bR^4=E_n\oplus E_m$ for $n,m\in \bZ$.  If $p: SU(2)\times \bR^4\rightarrow E_n\oplus E_{m}$ is the standard projection, then $\pi_* (L_3-n\partial_{\chi_1}-m \partial_{\chi_2})=0$ and so the vector fields induced by the left action
of $SU(2)$ on $SU(2)\times_{U(1)}\bR^4$ are
\bea
&&\xi_3=p_*R_3=\partial_\rho~,~~~\xi_2=p_*R_2=\sin(2\rho) \cot(2\eta) \partial_\rho-\cos(2\rho) \partial_\eta+ {\sin(2\rho)\over\sin(2\eta)} ( n\partial_{\chi_1}+m\partial_{\chi_2})~,
\cr
&&
\xi_1=p_*R_1=-\cos(2\rho) \cot(2\eta) \partial_\rho-\sin(2\rho) \partial_\eta-{\cos(2\rho)\over\sin(2\eta)} ( n\partial_{\chi_1}+m\partial_{\chi_2})~,
\eea
where we have set $y^1=r_1\cos\chi_1$, $y^2=r_1\sin\chi_1$,  $y^3=r_2\cos\chi_2$ and $y^4=r_2\sin\chi_2$.

To construct the most general invariant metric on $SU(2)\times_{U(1)}\bR^4$ first observe that the canonical connection $\Omega$ of $SU(2)/U(1)$ is $\Omega=s^*\tell^3=-\cos(2\eta) d\rho$, where the local section is chosen as $s(\eta, \rho)=(\eta, \rho, 0)$ as in the previous section.
The condition on $\Lambda$ in (\ref{ginvar2}) implies that $\Lambda=0$. Furthermore if in addition $n,m\not=\pm1$, (\ref{ginvar4}) also implies that $\Pi=0$. So the only twisting of the fibre coordinates of the fibration is induced by the canonical connection. Using this,  one finds that the invariant metric (\ref{invmetr}) in this case can be written as
\bea
&&ds^2=a^2 \delta_{AB} \ell^A \ell^B+ b_1^2\, dr_1^2 +b_2^2\, dr_2^2+b_3^2\,  (d\chi_1+  n s^*\tell^3)^2 +b^2_4\,(d\chi_2+  m s^*\tell^3)^2~,
\label{metrs22}
\eea
where  $a, b_1, b_2, b_3, b_4$ depend on the coordinates $ r_1, r_2$ and $m\chi_1-n\chi_2$, and $n,m\not=\pm1$.

The bundle $SU(2)\times_{U(1)}\bR^4$ is not a  neighbourhood of principal $S^2$ orbits for $n,m\not=0$. To see this observe that for $n, m\not=0$, the equation $\xi_r f=0$, $r=1,2,3$, has only three independent solutions $f=r_1, r_2, m\chi_1-n\chi_2$ instead of the four required for codimension four orbits. So the metric (\ref{metrs22}) describes the geometry of $M$ in a neighbourhood
of a special $S^2$ orbit of codimension 4.

\subsubsection{$S^3=SU(2)$ orbits}\label{prin}

As the isotropy group is the identity, $H=\{e\}$, both the representations on $\mathfrak{m}$ and $D$ on $V=\bR^k$ are trivial. As a result there is no contribution in the metric from either the canonical connection $\Omega$ or $\Lambda$.  However, the invariance condition (\ref{ginvar4}) is automatically satisfied and so
\bea
\Pi^a=\Pi^a_A \ell^A~,
\eea
where $\Pi^a_A$ are some constants and $\ell^A=\tell^A$, $A=1,2,3$, as given in (\ref{linvforms}).  In turn the invariant  metric (\ref{invmetr}) on $E=SU(2)\times \bR^k$ is
\bea
ds^2=g_{AB} \ell^A \ell^B+ \gamma_{ab} (dy^a+\Pi^a_A\, \ell^A)(dy^b+\Pi^b_B \, \ell^B)~,
\label{invmetrsu2x}
\eea
where now $g_{AB}$, $\gamma_{ab}$ and $\Pi^a_A$ depend on the coordinates $y$ of the fibre $\bR^k$. The Killing vector fields are $R_A$ given in (\ref{rightv}).
The metric (\ref{invmetrsu2x}) can admit a larger isometry group provided $g_{AB}$, $\gamma_{ab}$ and $\Pi^a_A$ are chosen appropriately, i.e. it can also
be invariant under the right action of $SU(2)$ generated by $L_A$ in (\ref{linvvf}).

To make a connection with the discussion on Kaluza-Klein theory below as well as to some global aspects of principal bundles, let us rewrite the metric (\ref{invmetrsu2x}) adapted to a fibration with fibre $SU(2)$, i.e. in the form (\ref{metr}). Indeed
\bea
ds^2=\hat g_{AB} (\ell^A+\hat \Pi^A_a dy^a) ( \ell^B+ \hat \Pi^B_b dy^b)+\hat \gamma_{ab} dy^a dy^b~.
\label{globsu2}
\eea
This is a local metric on  a principal $SU(2)$ fibration with fibre metric $\hat g_{AB}$  which depends on the base manifold coordinates $y$. Principal bundle theory is set up with the patching conditions to act   with left transformations on the typical fibe $SU(2)$. These do not commute with the isometries generated by $R_A$ and so the left group action generated by $R_A$ does not patch globally. As a result to retain globally an $SU(2)$ invariance, one has to require that the metric (\ref{globsu2}) is invariant under $SU(2)$
transformations acting on the right generated by $L_A$.  In turn this requires the conditions that
\bea
{\cal L}_{L_A} \lambda^B=\epsilon^B{}_{AC} \lambda^C~,~~~\hat g_{AD}\, \epsilon^D{}_{BC}+\hat g_{BD}\, \epsilon^D{}_{AC}=0~,
\eea
where
\bea
\lambda^A=\ell^A+\hat \Pi^A_a dy^a~,
\eea
is viewed as the principal bundle connection. Therefore the fibre metric $\hat g_{AB}$ must be bi-invariant but still can depend on the coordinates $y$.  If in addition the fibre metric $\hat g_{AB}$ is taken to be constant, then the metric (\ref{globsu2}) is the DeWitt
ansatz for a Kaluza-Klein vacuum with internal space the group manifold $SU(2)$ which always yields a consistent truncation of the Kaluza-Klein spectrum. This generalizes to all group manifold Kaluza-Klein reductions.

\subsubsection{$S^3=SU(2)\times_{U(1)} U(1)$ orbits}\label{prinsu22u1}

Next suppose that $SU(2)\times U(1)$ acts on a manifold with an  $S^3$ orbit.
 As a homogeneous space $S^3=SU(2)\times_{U(1)} U(1)$, where $U(1)$ acts on $SU(2)\times U(1)$ as $(g, u)\rightarrow (gv, v^{-1}u)$ generating the vector field $L^3$ on $SU(2)$.  Let $\pi: SU(2)\times U(1)\rightarrow SU(2)\times_{U(1)} U(1)=S^3$ be the standard projection.  Clearly $\pi_*(L_3+\partial_\psi)=\pi_*(-\partial_\tau+\partial_\psi)=0$, where $\psi$ is the standard angular coordinate on $\{e\}\times U(1)\subset SU(2)\times U(1)$.  The vector fields generated by the left action of $SU(2)\times U(1)$ on $ SU(2)\times_{U(1)} U(1)$ are given by
$\pi_*R_1, \pi_*R_2, \pi_*R_3, \pi_*\partial_\psi$,
 where $\pi_*R_1, \pi_*R_2, \pi_*R_3$  are expressed as in (\ref{rightv}) with $\partial_\tau$ replaced by $\partial_\psi$.

The vector field $L_3+\partial_\psi$ is generated by the Lie algebra elements $t^3+t^0\in \mathfrak{su}(2)\oplus \mathfrak{u}(1)$, where $t_0$ is the generator of $\mathfrak{u}(1)$. Therefore $t_3+t_0$ spans $\mathfrak{h}$. Choosing a splitting $\mathfrak{g}=\mathfrak{h}\oplus \mathfrak{m}$ with $\mathfrak{m}$ spanned
by $\{t_3- t_0, t_1, t_2\}$, the canonical principal bundle connection  $\theta\vert_{\mathfrak{h}}$ gives $\theta\vert_{\mathfrak{h}}(L_3+\partial_\psi)=t^3+t^0$ and $\theta\vert_{\mathfrak{h}}(X_{\mathfrak{m}})=0$, ie it vanishes
on all vector fields $X_{\mathfrak{m}}$ generated by the elements of $\mathfrak{m}$, where $\theta$ is the the Maurer-Cartan form on $SU(2)\times U(1)$.  A straightforward computation reveals that
\bea
\theta\vert_{\mathfrak{h}}={1\over2} (\hat\ell^3+\hat \ell^0)\, (t_3+t_0)~,
\eea
where $\hat\ell^3$ is given in (\ref{linvforms}) and $\hat \ell^0=d\psi$.
Choosing a local section $s: W\subset S^3\rightarrow SU(2)\times U(1)$, e.g.  $s(\rho, \eta, \tau)=(\rho, \eta, \tau, \tau)$, the canonical connection on the coset space is
\bea
\Omega&=&s^*\theta\vert_{\mathfrak{h}}={1\over2} s^*(\hat\ell^3+\hat \ell^0)\, (t_3+t_0)={1\over2} s^*(\hat\ell^3+d\psi)\, (t_3+t_0)
\cr &=&-{1\over2} \cos(2\eta)\, d\rho\, (t_3+t_0)~.
 \eea
 Note that the section $s$  is transversal to the integral curves $\tau+\psi=\mathrm{const}$, $\eta,\rho=\mathrm{const}$ of  $L_3+\partial_\psi$.

To continue suppose  that $S^3$ is a codimension 2 orbit and the associated vector bundle that models the neighbourhood of the orbit is
$E_n=(SU(2)\times U(1))\times_{U(1)} \bR^2$, where the diagonal $U(1)\subset SU(2)\times U(1)$ acts on $\bR^2$ with the $D_n$, $n\not=0$,  representation as in the previous examples.  In this case
$p_*(L_3+\partial_\psi-n\partial_\chi)=0$, where we have set $y^1=r\cos\chi, y^2=r\sin\chi$ for the coordinates of the fibre $\bR^2-\{0\}$.
The vector fields generated by the group action of $SU(2)\times U(1)$ on
$(SU(2)\times U(1))\times_{U(1)} \bR^2$ in a neighbourhood of the zero section are
\bea
\xi_1=p_*R_1~,~~~\xi_2= p_*R_2~,~~~\xi_3= p_*R_3~,~~~\xi_4= p_*\partial_\psi~,
\eea
where again $p_*R_1, p_*R_2, p_*R_3$ are given  as in
(\ref{rightv}) but now $\partial_\tau$ replaced with $\partial_\psi-n\partial_\chi$.   To construct the invariant metric, it follows from
(\ref{ginvar2}) that
\bea
\Lambda=f s^*(\tell^3-d\psi)~,
\eea
where $f$ is an arbitrary constant and  the section $s$ is again given by $s(\rho, \eta, \tau)=(\rho, \eta, \tau, \tau)$.  Furthermore for $n\not=\pm 1,0$, $\Pi=0$ as a consequence of (\ref{ginvar4}).  Using these
 the invariant metric on $(SU(2\times U(1))\times_{U(1)} \bR^2$ can be written as
\bea
ds^2=a_1^2 \delta_{A'B'} \ell^{A'} \ell^{B'}+ a_2^2 (\ell^3)^2+ b_1^2 dr^2+b_2^2 \left(d\chi+n\Omega+ nf s^*(\tell^3-d\psi)\right)^2~,
\label{su2u1metr}
\eea
where $a_1, a_2, b_1, b_2, f$ depend on $r$, $\ell^{A'}=s^*\tell^{A'}$,  $A',B'=1,2$,  $\Omega={1\over2} s^*(\tell^3+d\psi)=-{1\over2} \cos(2\eta) d\rho$ and $ \ell^3= {1\over2}  s^*(\tell^3-d\psi)=-{1\over2}\cos(2\eta) d\rho-d\tau$.

  As $U(1)$ bundles over $S^3$ are topologically trivial,  $(SU(2\times U(1))\times_{U(1)} \bR^2$ is topologically trivial. Nevertheless for $n\not=0$, it does not  model a neighbourhood of principal $S^3$ orbits. Indeed observe that $\xi_{\ur} h=0$, $\ur=1,2,3,4$, has one solution $h=r$ instead of two required for codimension 2 orbits.  Therefore, the metric (\ref{su2u1metr}) models
  the geometry around a special $S^3$ orbit.

  So to model the geometry of principal $SU(2)\times_{U(1)} U(1)$ orbits, one should take $n=0$. In such a case, one can demonstrate that
  \bea
  \Pi= e\, \ell^3~,
  \eea
  where $e$ is a constant.
  Then the invariant metric  reads
  \bea
ds^2=a_1^2 \delta_{A'B'} \ell^{A'} \ell^{B'}+ a_2^2 (\ell^3)^2+ b_1^2 dr^2+b_2^2 \left(d\chi +e\, \ell^3 \right)^2~,
\label{su2u1metrp}
\eea
  where now $a_1, a_2, b_1, b_2$ and $e$ are arbitrary functions of $\chi, r$. Note that although for principal orbits the representation $D$ is trivial, the metric
  above still contains rotation terms  and it is not just a warped product type of metric.

This example can be easily generalized to $SU(2)\times U(1)/U(1)_{p,q}$, where $p,q\in \bZ-\{0\}$ and co-prime.  In such a case $\mathfrak{h}$ is generated by $p t_3+q t_0$ which in turn generates the vector field $p L^3+q \partial_\psi$ on $SU(2)\times U(1)$. Choose   $\mathfrak{m}$  to be spanned by $\{p t_3-q t_0, t_1, t_2\}$.   Then the canonical connection is
\bea
\theta\vert_{\mathfrak{h}}={1\over2} ({1\over p}\hat\ell^3+{1\over q}\hat \ell^0)\, (p t_3+q t_0)~,
\eea
and so
\bea
\Omega={1\over2} s^*({1\over p}\hat\ell^3+{1\over q}\hat \ell^0)\, (p t_3+q t_0)~,
\eea
where $s$ is any local section of the fibration $U(1)\rightarrow SU(2)\times U(1)\rightarrow SU(2)\times U(1)/U(1)_{p,q}$.
Furthermore, one can demonstrate that
\bea
\Lambda=f s^*({1\over p}\tell^3-{1\over q}d\psi)~,
\eea
where $f$ is a constant.
Next consider invariant metrics on  $(SU(2)\times U(1))\times_{U(1)_{p,q}} \bR^2$.  For $n\not=\pm p$, one can show that $\Pi=0$ as a consequence of (\ref{ginvar4}). Using this the invariant metric
 on $(SU(2\times U(1))\times_{U(1)} \bR^2$ can be written as
\bea
ds^2=a_1^2 \delta_{A'B'} \ell^{A'} \ell^{B'}+ a_2^2 (\ell^3)^2+ b_1^2 dr^2+b_2^2 \left(d\chi+n\Omega+ nf s^*(\tell^3-d\psi)\right)^2~,
\label{su2u1metrpq}
\eea
where $a_1, a_2, b_1, b_2, f$ depend on $r$, $\ell^{A'}=s^*\tell^{A'}$,  $A',B'=1,2$,  $\Omega={1\over2} s^*({1\over p}\tell^3+{1\over q} d\psi)$ and $ \ell^3= {1\over2}  s^*({1\over p}\tell^3-{1\over q}d\psi)$.  Again the metric (\ref{su2u1metrpq}) for $p,q,n\not=0$ models the geometry in a neighbourhood of a special $SU(2)\times U(1)/U(1)_{p,q}$ orbit.

For principal orbits, one has to again take $n=0$.  After an analysis similar to that we have explained above, the invariant metric can be written as in (\ref{su2u1metrp}), where now $\ell^3={1\over2}  s^*({1\over p}\tell^3-{1\over q}d\psi)$.

\subsubsection{$S^3=SU(2)\times_{SU(2)} SU(2)$ orbits}\label{sec:su2}

As a final example consider $SU(2)\times SU(2)=\times^2SU(2)$ acting on a manifold with  $S^3=SU(2)\times_{SU(2)} SU(2)$ orbits. The right action of $SU(2)$ on $\times^2SU(2)$ generates the vector fields $L_\alpha-\tilde R_\alpha$, $\alpha=1,2,3$, where $L_\alpha$ are the left invariant vector fields given in (\ref{linvvf}) on $SU(2)\times \{e\}$ while $\tilde R_\alpha$ are the right-invariant vector fields given in (\ref{rightv}) on $\{e\}\times SU(2)$.  Therefore $L_\alpha-\tilde R_\alpha$ span the Lie algebra of the isotropy group $\mathfrak{h}=\mathfrak{su}(2)$. The $\times^2SU(2)$ action on $S^3=SU(2)\times_{SU(2)} SU(2)$ is generated
by $\pi_* R_r, \pi_* \tilde L_r$, $r=1,2,3$, where $\pi:\times^2SU(2)\rightarrow SU(2)\times_{SU(2)} SU(2)$.

Suppose now that the orbit $S^3$ has codimension   3.
As the principal bundle $\times^2SU(2)\rightarrow SU(2)\times_{SU(2)} SU(2)$ is topologically trivial,
all the associated vector bundles of this are topologically trivial and so the associated vector bundle  $E_D=\times^2SU(2)\times_{SU(2)}\bR^3$, where the representation
of $SU(2)$ on the typical fibre $\bR^3$ is the same as that of the isotropy group $SU(2)$ on $\mathfrak{m}$, i.e. $D$  is the standard vector representation.
The condition on $\Lambda$ in (\ref{ginvar2}) can be solved by setting $\Lambda= e {\bf 1}$, where $e$ a constant.  Similarly, the condition on $\Pi$ can be solved to yield $\Pi^a=f\delta^a_A s^*(\tell^A+\tilde \tr^A)$, where $f$ is a constant. As it is well known there is a single invariant metric on
$SU(2)\times_{SU(2)} SU(2)$ up to a constant.  Promoting these constants to invariant functions on the fibre $\bR^3$ under the action of the vector representation of $SU(2)$, one
finds that (\ref{invmetr})  reads
\bea
ds^2&=&b^2(r) \delta_{AB}\,(dy^A+(\Omega^\alpha + e \ell^D \delta_D^\alpha) f_{\alpha C}{}^A y^C+f \ell^A)(dy^B+(\Omega^\alpha + e(r) \ell^D \delta_D^\alpha) f_{\alpha C}{}^B y^C+f \ell^B)
\cr
&&\qquad\qquad
+a^2(r) \delta_{AB}\, \ell^A \,\ell^B
\cr
&=&b^2 \delta_{AB}\, (dy^A-(1+e) \ell^D \epsilon_{DC}{}^A y^C+f \ell^A)(dy^B-(1+e) \ell^D \epsilon_{DC}{}^B y^C+f \ell^B)
\cr
&&\qquad\qquad+a^2 \delta_{AB}\, \ell^A\, \ell^B~,
\label{s3s3}
\eea
where $a,b,e,f$ are functions of the radial coordinate $r$ of $\bR^3$ and we have used the local section $s:V\subset S^3\rightarrow \times^2 SU(2)$ with $s(g)=(g,e)$.  Thus $s^* \tilde \tr=0$ and so the connection is $\Omega^\alpha={1\over2}s^*(\tell^\alpha-\tilde \tr^\alpha)={1\over2}s^*\tell^\alpha$ and similarly the frame is $\ell^A={1\over2}s^*(\tell^A+\tilde \tr^A)={1\over2}s^*\tell^A$, where $\tell^A$ are the left-invariant forms (\ref{linvforms}) on $S^3$.

Before we complete the discussion notice that the most general $SO(3)$ invariant metric on $\bR^2$ is $b_1^2(r) dr^2+ b_2^2(r) r^2 ds^2(S^2)$ and so it is determined by two functions. However one of them can be eliminated using a coordinate transformation of $r$, e.g. set $b_1=b_2=b$.  This is in agreement with the form of
the metric in (\ref{s3s3}) in which the fibre metric depends on one function $b$.

Finally, let us test whether $S^3$ is a principal orbit in $E_D=\times^2SU(2)\times_{SU(2)}\bR^3$. For this, let us compute the isotropy group of the point
$(e,e,v)_D\in E_D$ for some $v\in \bR^3$, $v\not=0$. Recall the equivalence relation $(g_1 k, k^{-1} g_2, D(k^{-1} v)_D=(g_1,g_2, v)$, where
$g_1, g_2, k\in SU(2)$ and $v\in \bR^3$.  Acting with $(h_1, h_2)\in \times^2SU(2)$ on $(e,e,v)_D$ and demanding that $(h_1, h_2)$ is in the isotropy group of  $(e,e,v)_D$, one has that
\bea
(h_1, h_2) (e,e,v)_D=(h_1, h_2, v)_D= (h_1 k, k^{-1} h_2, D(k^{-1}) v)_D= (e,e,v)_D
\eea
which gives that $h_1=h_2^{-1}$ and $D(h_2) v=v$.  As $D$ is an orthogonal rotation and $v\not=0$, $h_2\in \{e\}\times U(1)\subset \times^2SU(2)$.  Thus the isotropy group of the nearby orbits to $N$ is $U(1)$ and therefore $N=SU(2)\times_{SU(2)} SU(2)$ is a special orbit.

\subsection{Applications to AdS backgrounds}\label{adsb}

Let us next turn to explore some applications in the context of supersymmetric AdS backgrounds. If   the R-symmetry is abelian, say $U(1)$, and acts (almost) effectively\footnote{Take the orbits to be closed.} on the internal space $M$ generating a vector field $X$, one can always adapt a coordinate $t$ to $X$, $X=\partial_t$. As $H=\{e\}$ and $G=U(1)$, the only contribution along the fibre in the invariant metric  (\ref{invmetr}) is from $\Pi$.  In particular the metric can be written as
\bea
ds^2= a^2 dt^2+\gamma_{ab} (dy^a+\Pi^a dt)(dy^b+\Pi^b dt)~,
\eea
where the only restriction on $a^2, \gamma, \Pi$ is that they should be independent of $t$ and otherwise depend on all $y$ coordinates. This metric has the form of (\ref{metrf2}) and it can be easily transformed to the familiar expression in (\ref{metr}).  In what follows, we shall focus on the geometry in the neighbourhood of principal obits.

\subsubsection{AdS$_6$ backgrounds}\label{secxx1}

  Since smooth IIB AdS$_6$ backgrounds with compact internal space are the only ones that have not been
  classified \cite{ads6}, let us apply our analysis above to this case. The Lie algebra of the R-symmetry group is $\mathfrak{so}(3)=\mathfrak{su}(2)$ and the only unresolved case is that with principal orbits  $S^2=SU(2)/U(1)$. The internal space of AdS$_6$ backgrounds in IIB has dimension 4,   the principal orbit $S^2$ has codimension 2 and  so the normal bundle has rank 2.  In addition from the results of section \ref{s2orbits2}, the normal bundle must be  associated with the trivial representation of the isotropy group $\mathfrak{u}(1)$ and so the metric of the internal space is given in equation (\ref{ps2metr}) but now for codimension 2 orbits. In particular, one finds that
  \bea
  ds^2=a^2(y) \delta_{AB} \ell^A \ell^B+ \gamma_{ab}(y) dy^a dy^b~,~~
  \label{anmetrx1}
\eea
which is the warped metric on $S^2\times \bR^2$ with $\gamma$ and $a^2$ an arbitrary metric and function  on $\bR^2$, respectively. As (\ref{anmetrx1}) is a local expression of the metric near a principal $S^2$ orbit,  $\bR^2$ can be taken as a chart in a 2-dimensional space $\Sigma$.  Thus the metric (\ref{anmetrx1}) is interpreted as a metric on the warped product $S^2\times \Sigma$. Incidentally, this is the ansatz used in the construction of  the IIB AdS$_6$ solutions in \cite{dhoker, dhoker2, dhoker3}  with non-compact internal space.

\subsubsection{AdS$_5$ backgrounds}

Next consider AdS$_5$ backgrounds. The maximally supersymmetric AdS$_5$ backgrounds have been classified in \cite{maxsusy} and those preserving 24 supersymmetries have been shown to be locally isometric to the maximally supersymmetric ones \cite{bgp}. AdS$_5$ backgrounds that preserve   16 and 8 supersymmetries are known to admit a $\mathfrak{u}(2)$ and $\mathfrak{u}(1)$ R-symmetry algebras, respectively. The latter  have already been dealt with as part of the general analysis of backgrounds with a $\mathfrak{u}(1)$ symmetry above.
It remains to consider the backgrounds preserving 16 supersymmetries. Up to discrete identifications, the homogeneous spaces which admit an (almost) effective $SU(2)\times U(1)$ action are $SU(2)\times U(1)$ and  $SU(2)\times U(1)/U(1)_{p,q}$, where $p,q\in \bZ$ are co-prime, $p\not=0$.

In what follows, let us seek metrics on  the internal spaces for AdS$_5$ backgrounds with  $SU(2)\times U(1)$ and  $SU(2)\times U(1)/U(1)_{p,q}$ as principal orbits.
In a type II theory in 10 dimensions, the principal orbit  $SU(2)\times U(1)$ has   codimension 1 in the internal space.  So the most general metric invariant metric on the internal space is
\bea
ds^2=g_{AB} \ell^A\ell^B+ b^2 (dy+ \Pi_A \ell^A)(dy+ \Pi_B \ell^B)~,
\label{metrsu2u1xx}
\eea
where the metric $g_{AB}$ on $SU(2)\times U(1)$, $\Pi$ and $b^2$  depend only on $y$. The metric above can be written as a principal bundle metric.  The results in section \ref{prin} obtained
for $SU(2)$ can be easily adapted for $SU(2)\times U(1)$.

Next turn to investigate the internal spaces with principal $SU(2)\times U(1)/U(1)_{p,q}$ orbits. These in type II 10-dimensional theories have codimension 2 in the internal space. A detailed analysis has already been carried out in section \ref{prinsu22u1}.  The metric on the internal space is given in
 (\ref{su2u1metrp}) for manifolds with a $SU(2)\times U(1)/U(1)$ orbit, $p=q=1$. For the rest it is again given in (\ref{su2u1metrp}) after an appropriate definition of $\ell^3$, see discussion in section \ref{prinsu22u1}.

\subsubsection{AdS$_4$ backgrounds}

Next consider AdS$_4$ backgrounds. The internal spaces of those preserving 4 supersymmetries admit no R-symmetries and those preserving 8 supersymmetries
admit an $\mathfrak{so}(2)$ symmetry that we have already investigated.  Backgrounds preserving 12 supersymmetries admit an $\mathfrak{so}(3)=\mathfrak{su}(2)$ action. There are two kind of orbits, $SU(2)$ and  $SU(2)/U(1)=S^2$,  up to discrete identifications,  that can occur admitting an (almost) effective $SU(2)$ action.  We have already described   manifolds with principal $SU(2)$ and  $SU(2)/U(1)=S^2$ orbits in the examples above, see sections \ref{prin} and  \ref{s2orbits2}.

It remains to investigate AdS$_4$ backgrounds that preserve 16 supersymmetries. These must admit an (almost) effective $\mathfrak{so}(4)$ action. The homogeneous spaces which admit an (almost) effective $SO(4)$ action, up to discrete identifications,  are
\bea
&&SO(4)~,~~~SO(4)/SO(2)_{m,n}~,~~~~SO(4)/(SO(2)\times SO(2))=S^2\times S^2~,~~~
\cr
&&SO(4)/SO(3)=S^3~,
\eea
where $m, n$ are integers, (relatively prime), which specify the embedding of $SO(2)$ into $SO(4)$.

Let us begin with principal $SO(4)$ orbits. Such orbits are of codimension 0 in  the internal spaces of 10-dimensional backgrounds and of codimension 1 in 11-dimensional backgrounds.  In the former case, the internal manifold is homogeneous.  In the latter case, the metric on the internal space can be written as in
(\ref{metrsu2u1xx}) and the description of the components is the same as in the previous case but now $SU(2)\times U(1)$ is replaced by $SO(4)$, see also section \ref{prin}.

Next let us explore internal spaces of AdS$_4$ backgrounds with principal $SO(4)/SO(3)=SU(2)\times_{SU(2)}SU(2)=S^3$ orbits in type II 10-dimensional theories.
These have codimension 3 in the internal space. It is clear that from the results of section \ref{sec:su2} that the representation $D$ of the little group $SU(2)$
on $\bR^3$ must trivial.  It is then straightforward to verify that $\Pi=0$. The invariant metric on the internal space can be written as
\bea
ds^2=b^2 \delta_{AB}\, \ell^A \ell^B+ \gamma_{ab}\, dy^a dy^b~,
\eea
where $b^2$ and $\gamma$ depend only on the coordinates $y$. The metric $g_{AB}=b^2 \delta_{AB}$ is required as the metric on $S^3$ must be both left- and right-invariant.

From the remaining cases, let us consider internal spaces with principal $SO(4)/SO(2)_{m,n}$, $m,n\not=0$,  orbits. These are of codimension 1 in the internal spaces
of 10-dimensional AdS$_4$ backgrounds.  The $D$ representation of $SO(2)$ on the fibre is trivial, therefore the canonical connection and $\Lambda$ do not contribute to the metric. Provided that the isotropy group $SO(2)$ generates the vector field $mL^3-n\tilde{\hat R}^3$ on $SO(4)\sim SU(2)\times SU(2)$, the conditions (\ref{ginvar4}) imply that
\bea
\Pi= {1\over2} e \, s^* \big({1\over m} \hat\ell^3+{1\over n} \hat{\tilde r}^3\big)~,
\eea
where $e$ is a constant and $s$ a local section.
After solving the invariance conditions (\ref{ginvar1}), one finds that the  invariant metric on the internal space is
\bea
ds^2=a_1^2 \delta_{A'B'} \ell^{A'}\ell^{B'}+b_1^2 \delta_{\tilde \tilde A' \tilde B'} r^{\tilde A'} r^{\tilde B'}+ a_1^2 (\ell^3)^2+ a_2^2 (dy+ e \ell^3)^2~,
\eea
where $\ell^{A'} =s^*\tell^{A'}$, $r^{\tilde A'} =s^*r^{\tilde A'}$, $A', \tilde A'=1,2$,  $\ell^3={1\over2}  s^* \big({1\over m} \hat\ell^3+{1\over n} \hat{\tilde r}^3\big)$, and $a_1, a_2, b_1, b_2, e$ depend on $y$.
  A more systematic investigation of  AdS backgrounds which will include the remaining invariant fields of the associated  theories will be presented elsewhere.

\section{Concluding Remarks}

We have provided a systematic way to construct  invariant metrics and other invariant fields under the action of a (non-abelian) group $G$ on a manifold $M$.
Such metrics model the invariant geometry around an orbit $N$ in $M$ of the group $G$ either this orbit is principal or special.  For this we utilized the geometry of homogeneous spaces, $N=G/H$,  together with the slice and principal orbit theorems. The slice theorem provides a local model of the action of a group in a neighbourhood around an orbit, e.g. provides an expression for the vector fields generated by the group action in a convenient coordinate system.   We presented several examples that
illustrate the construction mostly focused on $S^2$ and $S^3$ orbits.  The main results are given in equations (\ref{invmetr}) and (\ref{informsx1}) for the invariant metrics and forms, respectively. These expressions include all the ansatzes used in the literature to describe such invariant fields  under the action of a group. Furthermore, we used our results to construct invariant metrics on the internal space of AdS backgrounds under the action of the R-symmetry group with main focus on the geometry in  a neighbourhood of a principal orbit.

As the invariant metric  (\ref{invmetr}) provides a model for the local geometry of $M$ around any orbit,  either the orbit is  principal or special, (\ref{invmetr}) can also be used to investigate the geometry of the internal spaces of AdS backgrounds that contain special orbits of an R-symmetry group. Combined this with the results we have described in section \ref{adsb} will provide a complete description of the local geometry  of  internal spaces. Of course, the expression for the metric in
(\ref{invmetr}) solves the kinematic problem.  To find a background one also has to solve the field equations of a theory. Nevertheless, the approach proposed is systematic and the problem is further simplified for the supersymmetric backgrounds.

The metric (\ref{invmetr}) can also be used  to describe  Kaluza-Klein ansatzes with  internal spaces $M$ that admit the action of an isometry group $G$.  For this, one first rewrites the metric (\ref{invmetr}) on $M$   as a fibration with fibre the orbit $N=G/H$ and then changes coordinates to Frobenius coordinates, i.e. rewrite the metric (\ref{invmetr}) as in (\ref{metr}). As the vector fields $\xi_{\ur}$ are isometries, one can gauge these isometries by adding a Pauli term, i.e. replace $dw^I$ with $dw^I-A_m^{\ur} \xi_{\ur}^I dz^m$ in (\ref{metr}), where $A$ is a gauge field that depends on the coordinates of the lower dimensional spacetime. Furthermore, one can allow the various parameters that determine (\ref{invmetr}) to depend on the lower-dimensional spacetime coordinates.  These can be thought as the breathing modes.  Several simplifications may be possible after a careful selection of the allowed breathing modes. Of course it  is not apparent that such an ansatz   will lead to consistent truncation to a lower dimensional theory, see e.g. \cite{pope} for a recent discussion and references within.  Nevertheless as (\ref{invmetr}) can describe the geometry in the vicinity of special orbits, one  can   model a Kaluza-Klein scenario,  where
the principal orbit $N=G/H$ in the internal space $M$  of a compactification degenerates to a  special orbit while  the number of gauge fields $A$ of the lower dimensional theory  remain the same.

\section*{Acknowledgments}

I would like to thank Jurgen Brendt for  helpful discussions.

\end{document}